\documentclass[fleqn,usenatbib,usedcolumn]{mnras}

\usepackage{newtxtext,newtxmath}
\usepackage[T1]{fontenc}
\usepackage[british]{babel}
\usepackage{ae,aecompl}

\usepackage{graphicx}
\usepackage{amsmath}
\usepackage{amssymb}

\usepackage{subcaption}
\title[UHE cosmic rays from nearby starburst galaxies]{Ultrahigh energy cosmic rays from nearby starburst galaxies}

\author[R. Attallah and D. Bouchachi]{Reda Attallah\thanks{E-mail: reda.attallah@gmail.com} and Dallel Bouchachi \\
Department of Physics,  Faculty of Science, Badji Mokhtar University, P. O. Box 12, Annaba 23000, Algeria}

\date{Accepted XXX. Received YYY; in original form ZZZ}
\pubyear{2016}

\begin{document}
\label{firstpage}
\pagerange{\pageref{firstpage}--\pageref{lastpage}}
\maketitle

\begin{abstract}
Ultrahigh energy cosmic rays are the most energetic of any subatomic particles ever observed in nature. The quest for their mysterious origin is currently a major scientific challenge. Here we explore the possibility that these particles originate from nearby starburst galaxies, a scenario that matches the recent observation by the Telescope Array experiment of a cosmic-ray hotspot above 57~EeV not far from the direction of the starburst galaxy M82. Specifically, we study the stochastic propagation in space of ultrahigh energy cosmic rays through the state-of-the-art simulation framework CRPropa~3, taking into account all relevant particle interactions as well as deflections by the intervening magnetic fields. To ensure a comprehensive understanding of this model, we consider the energy spectrum, the cosmogenic neutrinos and gamma rays, and the distribution of arrival directions. The starburst galaxy scenario reproduces well observations from both the Telescope Array and Pierre Auger Observatories, making it very attractive for explaining the origin of cosmic rays at the highest energies.
\end{abstract}


\section{Introduction}
\label{sec:introduction}
Ultrahigh energy (UHE) cosmic rays are the highest-energy particles arriving on Earth from outer space. With energies exceeding 1~EeV ($\equiv 10^{18}$~eV), they are obviously accelerated by the most violent phenomena in the Universe. Although observational evidence strongly suggests an extragalactic origin, no source has clearly been identified. Nor has the mechanism by which they acquire such incredibly high energies. Literatures suggest active galactic nuclei (AGN), gamma-ray bursts (GRBs), magnetars, pulsars, galaxy clusters, starburst galaxies~\ldots, but no explanation is yet conclusive. Solving the puzzle of UHE cosmic rays is of paramount importance in modern astrophysics \citep[and references therein]{KOT11, LET11}. 

As UHE cosmic rays propagate from their sources to Earth, they interact with the intergalactic medium. Thus, they suffer a certain number of energy losses that affect the initial energy spectrum and mass composition. The most important of these processes involves the interaction with the low-energy photons of the cosmic microwave background (CMB). For protons, this phenomenon, called pion photoproduction, occurs above a threshold energy of $\sim 60$~EeV \citep{GRE66, ZAT66}. In other words, the energies of cosmic-ray protons should then not exceed this limit, known as the GZK cutoff, if they are cosmological in origin.  The photo-disintegration of heavy cosmic-ray nuclei would have a similar effect \citep{ALL05}. The actual observation of cosmic rays above the GZK cutoff makes it clear that these particles come from some nearby sources in the local Universe, which have yet to be identified.

Considerable progress in this field has been made in recent years with the advent of a new generation of large scale experiments, namely the Pierre Auger Observatory near Malarg\"{u}e (Argentina) \citep{AAB15NIM} and the Telescope Array (TA) experiment in Utah (USA) \citep{ABU12, TOK12}. Both are hybrid detectors, employing two complementary observation techniques. They combine a huge array of surface detectors with a collection of large air fluorescence telescopes. This experimental strategy improves the quality and quantity of data significantly. The main experimental observables are the all-particle energy spectrum, the shower maximum depth distribution which is very sensitive to the mass of the primary cosmic-ray particle, and the arrival direction distribution.

The observed energy spectrum exhibits above 1~EeV two major features not yet fully understood. There is first a flattening at $\sim$~4~EeV known as the ``ankle'' of the spectrum, and then a flux suppression above $\sim$~40~EeV \citep{ABB08, ABR10, ABU13ApJL}. Although the suppression of the flux occurs at almost the same energy as expected from the GZK effect, one cannot exclude the possibility of a running out of energy at the cosmic-ray accelerators \citep{ALO11}.

The energy spectrum cannot alone make it possible to conclude about the origin of UHE cosmic rays. The study of the mass composition offers additional key information and provides stringent constraints on acceleration and propagation models. Unfortunately, the determination of the primary mass in extensive air shower (EAS) measurements is not possible on an event-by-event basis. It is rather inferred on a statistical basis from the comparison of experimental observables, usually the shower maximum depth, with predictions from theoretical models. The hadronic interaction, which is poorly known at UHE, entails a number of uncertainties in EAS simulations and hence in the assessment of the mass composition. The present experimental situation is still confusing. The Auger data indicate a trend towards a heavy elemental composition above 10~EeV \citep{AAB14PRD}, whereas the TA data suggest a proton predominance in the same energy range \citep{ABB15}. But the uncertainties are still too large to be conclusive. In fact, a joint working group from both collaborations addressed this issue and found that the two results are quite consistent within systematic uncertainties \citep{UNG15}.

In theory, UHE cosmic rays cannot be confined by diffusive propagation within our galaxy and should then show significant anisotropy in their arrival directions. But because of the extreme weakness of their flux, the search for anisotropy has proved very difficult. The Auger data do not show evidence of deviation from isotropy, with the exception of a ``warm spot'' above 58~EeV around the direction of the nearby radio galaxy Centaurus A \citep{AAB15ApJ}. However, the TA data indicate a statistically significant cosmic-ray ``hotspot'' above 57~EeV in the northern sky near the Big Dipper \citep{ABB14}. Several studies have shown that the distribution of the TA hotspot events is consistent with the hypothesis of a single source, the nearby starburst galaxy M82 being the most promising candidate \citep{PFE17, HE16, FAN14}.

Experimental data on UHE cosmic rays are typically interpreted in the context of propagation models and a number of numerical simulations have been developed to this end (see, e.g., \citet{HAC16, TAK12, STA09, HOO07}). However, there is not yet a definitive scenario as aforementioned. Here we present the results of our extensive numerical investigations of the propagation of UHE cosmic rays in space, carried out with the help of the publicly available software CRPropa version~3 \citep{BAT16}. This computer code simulates the galactic and extragalactic propagation of UHE cosmic-ray nuclei and their secondary particles taking into account all relevant particle interactions and deflections by magnetic fields. We explore in particular the possibility that UHE cosmic rays originate from nearby starburst galaxies, which is supported by the TA observation of a hotspot not far from the direction of the starburst galaxy M82. In addition to experiencing an exceptionally high star formation rate (SFR), the active regions of starburst galaxies often contain large amounts of very dense molecular gas and host intense magnetic and radiation fields \citep{MUX06}. These extreme conditions make them a preferred environment for accelerating cosmic-ray particles to the highest energies \citep{ANC18, YOA15, ACC09, ACE09, ANC01, ANC99}. To ensure a comprehensive understanding of the starburst galaxy scenario, we have conducted a combined investigation of the all-particle energy spectrum, the cosmogenic neutrinos and gamma rays, and the anisotropy in the arrival directions.

\section{Propagation and source models}
\label{sec:models}
CRPropa~3 implements all relevant interactions of UHE cosmic-ray nuclei: pion photoproduction, pair production and photo-disintegration with the cosmic microwave background (CMB) and the UV/optical/infrared background radiations as well as nuclear decay \citep{BAT16}. The secondary particles created in these interactions are also included. The pion photoproduction is handled by the SOPHIA software \citep{MUC00}.

CRPropa~3 offers different propagation modes. The one-dimensional (1D) mode allows to take into consideration the cosmological evolution of sources and background radiations as well as the adiabatic energy losses. The three-dimensional (3D) mode allows for the distribution of sources and the magnetic fields to be defined on a 3D grid, thus enabling us to perform simulations in more realistic scenarios. The observed properties of UHE cosmic rays depend to a large extent on the galactic magnetic field (GMF) and the intergalactic magnetic field (IGMF).

The key quantities of the UHE cosmic-ray sources are the injection energy spectrum, the maximum acceleration energy, the initial mass composition and the large scale distribution. Since the observed energy spectra of several cosmic-ray elements are well described by an inverse power law in energy, it is natural to expect an injection spectrum of the same shape but with an exponential cutoff:
\begin{equation}
  \frac{\mathrm{d} N}{\mathrm{d} E} \propto E^{-\gamma}  \mathrm{e}^{-E/E_\mathrm{cut}},
  \label{eq:energy}
\end{equation}
\noindent where $E$ is the energy, $N$ is the number of cosmic-ray particles, $\gamma$ is the spectral index, and $E_\mathrm{cut}$ is the energy cutoff. The acceleration of cosmic rays in astrophysical shock fronts (Fermi mechanism) leads to spectral index values roughly in the range 2-2.5 (see, e.g., \citet{BEL78}). In extreme cases, such as found in ultra-relativistic shocks, the spectral index could however be as hard as 1 \citep{SCH15}. The potential sources of UHE cosmic rays, which include active galactic nuclei, gamma-ray bursts, neutron stars~\ldots, are capable of accelerating charged particles up to energies of about $10^{20.5}$~eV $\approx 300$~EeV \citep{HIL84}. In fact, the maximum energy attainable by a cosmic-ray accelerator depends heavily on its size and the strength of its magnetic field. It also depends on the nature of the particle to be accelerated, accelerating protons being more difficult than accelerating heavier nuclei.

In this work we used CRPropa~3 with all settings set to default. We used the catalog of the brightest nearby starburst galaxies (with redshifts $z < 0.03$) drawn up by \citet{BEC09}. We assumed that the intensity of UHE cosmic rays scales directly with the star formation activity \citep{BIE16, ACC09}. Then we simply weighted each source by its relative far-infrared (FIR) luminosity observed at 60~$\mu$m, which is a good indicator of the SFR \citep{MUX06}. Since M82 is the brightest FIR starburst galaxy, this assumption enables us to automatically reproduce the TA hotspot. But even more interesting is that the Auger warm spot is also reproduced in this way (see Section~\ref{subsec:anisotropy}). We considered two different initial mass compositions: the first (pure proton) matching the TA observations, and the second (mixed) matching the Auger observations. We assumed the mixed composition to comprise H, He-4, N-14, Al-27 and Fe-56 nuclei with equal abundances, corresponding to a metallicity $\sim$ 10 times that of the typical galactic cosmic-ray component \citep{ENG90}. To ensure sufficient statistics at high energy, we first injected particles with a power law $\propto E^{-1}$ up to the maximum energy which is deemed charge dependent, and reweighed afterwards in order to comply with Eq.~\ref{eq:energy} \citep{VLI14}. At the same time we tuned up the injection parameters ($\gamma$ and $E_\mathrm{cut}$) so as to achieve the best fit to the observed data.

\section{Results}
\label{sec:results}

\begin{figure} \centering
  \includegraphics[width=1.0\columnwidth]{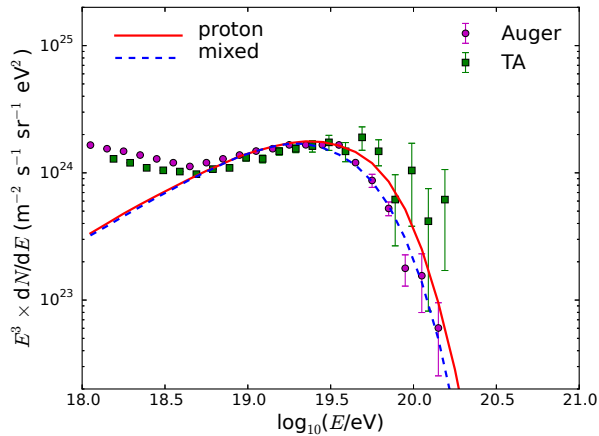}
  \caption{All-particle energy spectra of UHE cosmic rays from nearby starburst galaxies for the two mass compositions (Auger normalization at 10~EeV). Also shown are the experimental data from Auger \citep{VAL15} and TA \citep{IVA15}. The TA energy scale is shifted down by 13\% \citep{UNG15}.}
  \label{fig:spectrum}
\end{figure}

\subsection{All-particle energy spectrum}
\label{subsec:spectrum}
The all-particle energy spectrum is the most significant observable in cosmic-ray physics. It provides valuable information about the sources of cosmic rays as well as the media in which they propagate. Reproducing as faithfully as possible the striking features of the energy spectrum is a key point of any scenario for the origin of cosmic rays.

As the first step in our investigation, we considered the 1D-propagation of $10^6$ UHE cosmic-ray particles and calculated the all-particle energy spectrum for the two initial mass compositions (Fig.~\ref{fig:spectrum}). We obtained good agreement with observations for $\gamma = 2.2$ and $E_\mathrm{cut} = 30$~EeV in both cases. The spectrum is normalized to the Auger flux at 10~EeV and the TA energy scale is shifted down by 13\%. This shift has been advocated by the Auger-TA Composition Working Group after comparing the energy spectra of the two experiments at the ankle region \citep{UNG15}. The model does not, however, reproduce the ankle structure and can account for the energy spectrum only at the highest energy. This deficiency, which is not limited to this model, can be addressed with recourse to a second component of galactic cosmic rays \citep{HIL06}, produced for instance by supernova explosions of Wolf-Rayet stars \citep{THO16}. The ankle would then reflect, as is generally believed, the transition from the steepening galactic component to the hardening extragalactic one. On the other hand, the flux suppression observed here does not reflect the GZK effect but rather the depletion of the accelerator power. 

It should be noted that this result is not specific to starburst galaxies. Any type of sources with similar injection spectrum and composition can result in such a fit to the Auger and TA data above the ankle, as shown by many previous studies (see, e.g., \citet{KAM14, KOT11}). Furthermore, it is worth mentioning that the well-known ``dip model'' reproduces the ankle of the energy spectrum without the need for a Galactic component \citep{BER06}. This alternative model is based on the premises that UHE cosmic-ray sources are cosmological in origin and that the mass composition at injection is proton-dominated. The latter assumption fits better with TA data than Auger. Anyway, the issue remains under discussion and there is nothing conclusive yet.

\subsection{Cosmogenic neutrinos and gamma rays}
\label{subsec:cosmogenic}
As mentioned previously, while propagating through space UHE cosmic-ray protons interact with the CMB photons above $\sim 60$~EeV and produce pions. Pions are unstable particles and decay very rapidly, thus yielding the so-called cosmogenic neutrinos and gamma rays. High-energy neutrinos are also created in the decay of muons and neutrons, other products of this phenomenon. The interaction with background radiations of heavier UHE cosmic-ray nuclei also produces high-energy neutrinos and gamma rays, but in significantly smaller quantities. These secondary messengers are of great interest because they can be used to test the different models of the origin of UHE cosmic rays on an independent basis. Cosmogenic particles have not been observed so far, but nevertheless upper limits on the absolute fluxes of UHE primary cosmic-ray photons and neutrinos have been derived from the analysis of experimental data \citep{AAB17, AAB15PRD, BLE15, RUB17, AAR16}.

The calculation of cosmogenic particles must not just rely on nearby starburst galaxies, as they can reach Earth even from cosmological distances (especially neutrinos). Therefore, one should take into account the secondary production of the starburst galaxy population from all redshifts. Since it is practically impossible here to weight the sources individually, we considered the simple case of a homogeneous distribution of identical sources in a 1D-simulation of $10^6$ cosmic-ray particles. We assumed that sources evolve as the SFR whose function is of the form $(1+z)^m$, where $z$ is the source redshift and $m$ the evolution parameter. The latter is as follows \citep{YUK08}:
\begin{equation}
    m = \left\{
    \begin{array}{llc}
	3.4  & \mathrm{for} & z \le 1 \\
	-0.3 & \mathrm{for} & 1 < z \le 4 \\
	-3.5 & \mathrm{for} & z > 4.
    \end{array} \right.
\end{equation}
\noindent We fixed the maximum value of $z$ at 6. As before, the all-particle energy spectrum is normalized to the Auger flux at 10~EeV and the resulting cosmogenic-particle fluxes are normalized accordingly. We selected the energies of the primary particles from the same power law that leads to the best match with the observed energy spectrum (cf. Section~\ref{subsec:spectrum}). Cosmogenic neutrinos move along straight paths and are subject only to adiabatic energy loss. Cosmogenic gamma-rays together with the high-energy electrons also produced in these interactions form electromagnetic cascades. We considered the development of the electromagnetic cascades as part of the CRPropa simulation chain \citep{HEI18}.

\begin{figure} \centering
  \includegraphics[width=0.94\columnwidth]{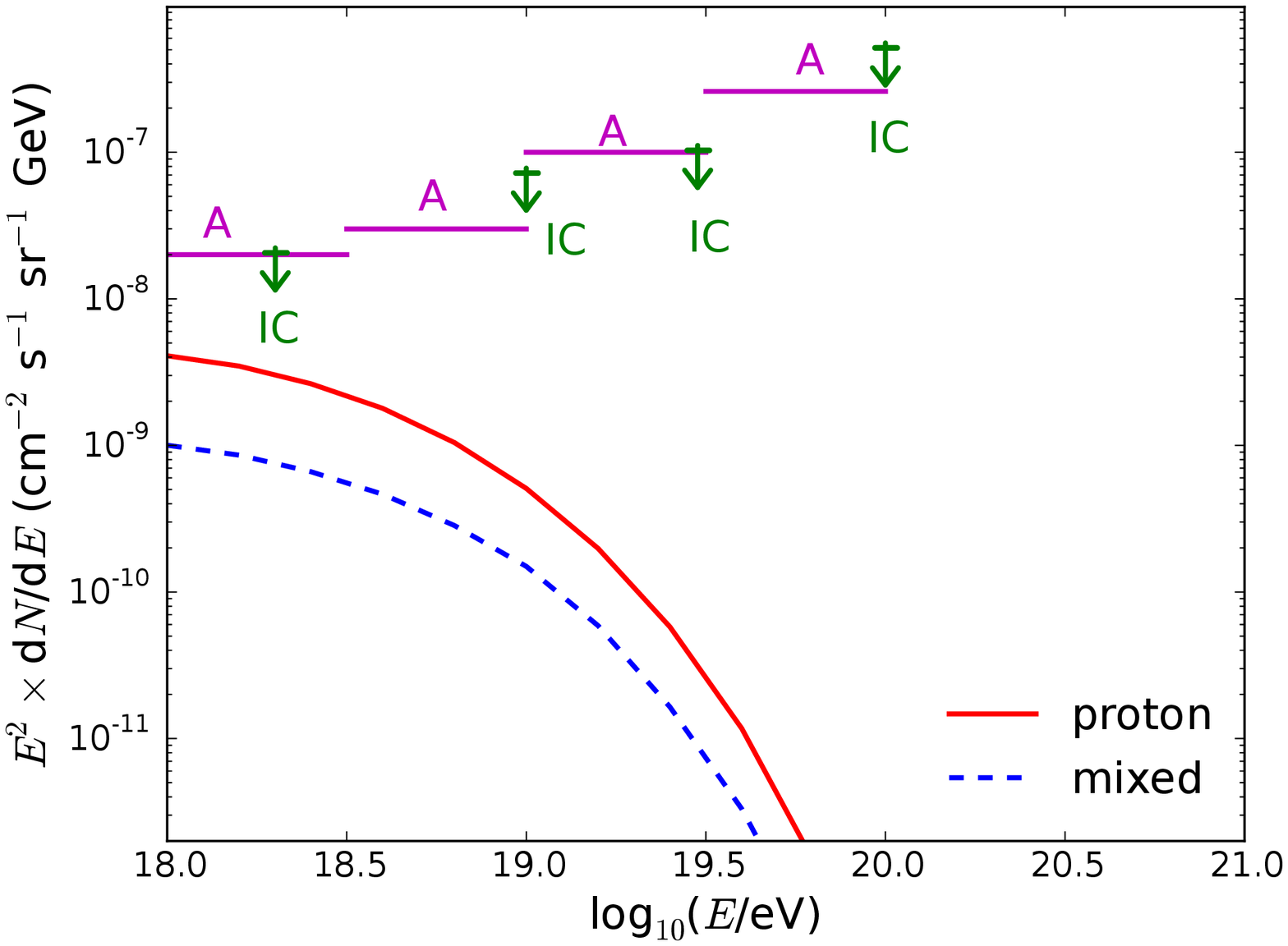}
  \includegraphics[width=0.94\columnwidth]{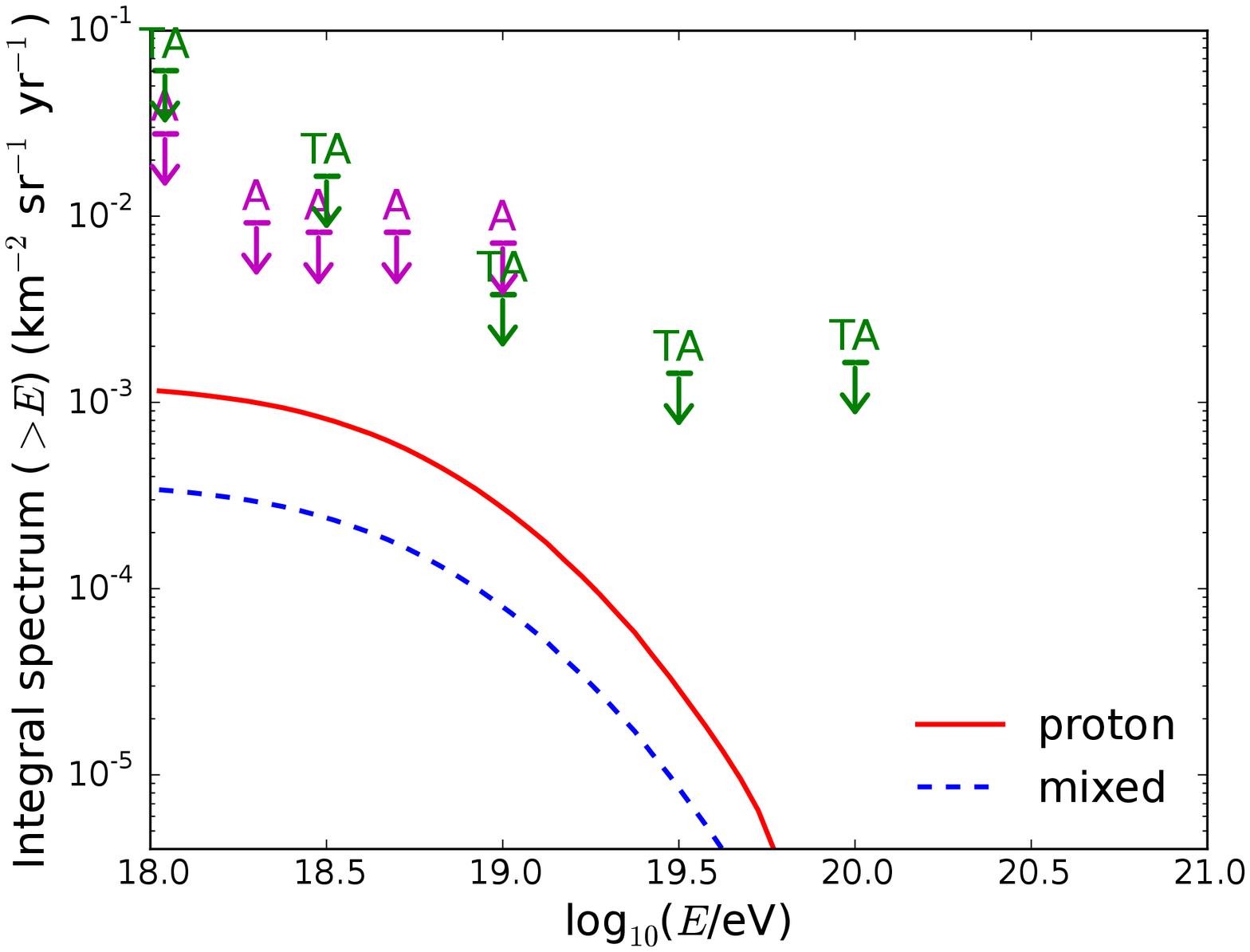}
  \caption{Energy spectra of cosmogenic neutrinos (top) and cosmogenic gamma-rays (bottom) produced by UHE cosmic rays from starburst galaxies for the two initial mass compositions. Also shown are the upper limits to the flux of UHE cosmic-ray photons and single-flavor neutrinos set by Auger (A) \citep{AAB17, AAB15PRD, BLE15}, TA (TA) \citep{RUB17} and IceCube (IC) \citep{AAR16}.}
  \label{fig:cosmogenic}
\end{figure}

It can be seen from Fig.~\ref{fig:cosmogenic} that the energy spectra of cosmogenic neutrinos and gamma rays from starburst galaxies are lower by several orders of magnitude than the upper bounds set by the different experiments. As could be expected, the starburst galaxy scenario is not challenged by the test of cosmogenic particles. But here again, this result is not unique to starburst galaxies. Any type of sources in similar conditions can produce fluxes of cosmogenic particles well below current instrument sensitivities, as shown by many previous studies (see, e.g., \citet{VLI17, ALO15, STA14, KAM12, KOT10}).

\subsection{Anisotropy}
\label{subsec:anisotropy}
The assessment of the degree of anisotropy in the arrival directions of UHE cosmic rays obviously calls for 3D-simulation. That is to say, it is necessary to consider also the spatial distribution of sources and the deflections caused by the intervening magnetic fields (GMF and IGMF). Our understanding of the GMF has improved significantly in recent years and seve\-ral successful models are now available. In contrast, the IGMF is poorly known and the results of theoretical studies do not corroborate, predicting deflection angles for UHE cosmic rays from sources in the local Universe over a rather wide range \citep{SIG04, DOL05, DAS08, BAT17}. Commonly, the IGMF is thought to be random with a correlation length $\lesssim 1$~Mpc and a magnitude $\lesssim 1$~nG \citep{PSH16}. As such, the effects of the IGMF would be simply to spread a signal around the direction towards the source. In CRPropa~3 the GMF and the IGMF are set up in separate simulations \citep{BAT16}. The first simulation deals solely with the propagation through the IGMF from sources to the edges of our galaxy, and then the second simulation takes over and addresses the propagation in the GMF.

In this work we used the full JF12 GMF model \citep{JAN12ApJ, JAN12ApJL} through the interface provided by CRPropa~3. To efficiently account for the effects of deflection in the GMF, we employed the ``lensing technique'', initially developed for the PARSEC software \citep{BRE14}. We assumed the IGMF simply as a random Kolmogorov-like turbulent field with a power spectrum on 60-1000~kpc length scales, which corresponds to a correlation length of $\sim 230$~kpc, and an RMS field strength of 1~nG. We first tracked $10^3$ primary particles ($E>57$~EeV) up to the edges of a sphere of 20~kpc radius representing the Milky Way, and then we applied the galactic lens. Needless to say, the overwhelming majority of simulated cosmic rays miss the detection sphere; we had to simulate $\sim$~2$\times$10$^{8}$ particles to end up with $10^3$ particles reaching our galaxy. We analyzed the resulting distribution of arrival directions with the standard software HEALPix \citep{GOR05}. For all our sky maps we used a grid of 49,152 pixels, the default angular resolution of CRPropa~3. To enhance sensitivity to week anisotropic signals, we utilized oversampling with an angular scale of $20^\circ$ \citep{HAY99}. In addition, we calculated the statistical significance $S$ of the excess at each pixel using the standard Li-Ma formula \citep{LI83}:
\begin{equation}
  \frac{S^2}{2} = N_\mathrm{on} \ln \left[ \frac{(1+\eta)N_\mathrm{on}}{\eta (N_\mathrm{on}+N_\mathrm{off})} \right] +
		  N_\mathrm{off} \ln \left[ \frac{(1+\eta)N_\mathrm{off}}{N_\mathrm{on}+N_\mathrm{off}} \right].
\end{equation}
\noindent $N_\mathrm{on}$ is simply the content of the pixel (after oversampling). To estimate the number of background events under the signal in $N_\mathrm{on}$, we generated $10^4$ events assuming a well-isotropic flux. The content of each pixel is then overwritten as before with the cumulative value of all the surrounding pixel contents up to 20$^\circ$. The obtained value is defined as $N_\mathrm{off}$. The normalization factor $\eta$ is the ratio of the total number of pseudo-data ($10^3$ in our case) to the total number of isotropic events ($10^4$), that is $\eta = 0.1$. This way, the background in each pixel is just $N_\mathrm{bg} = \eta N_\mathrm{off}$. $S$ provides an assessment of the statistical significance of the signal deviation with respect to a well-isotropic flux. We should underline here that we did not take into account any special experimental exposure in the calculation of $S$.

\begin{figure} \centering
  \includegraphics*[width=0.8\columnwidth]{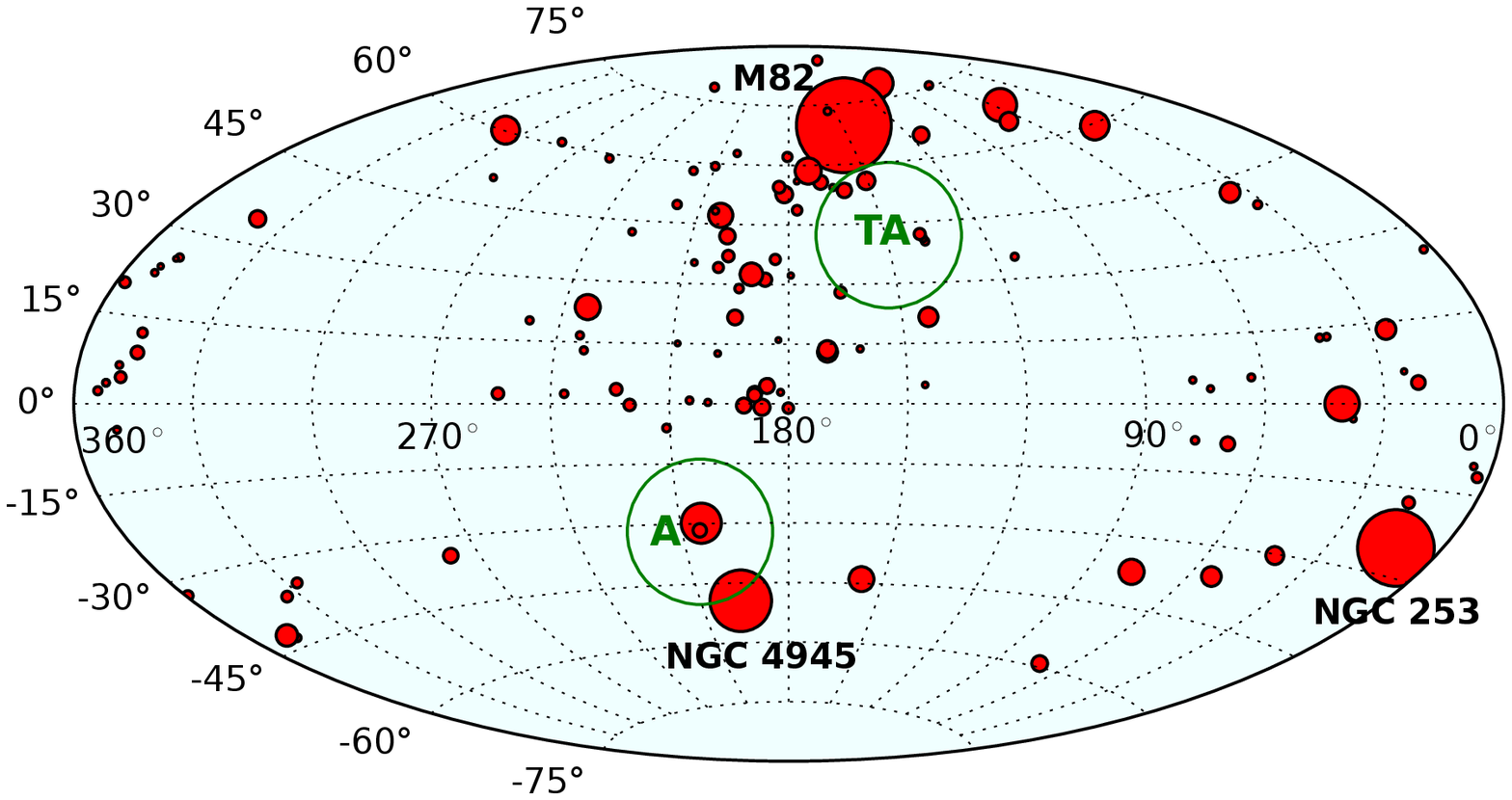}
  \includegraphics*[width=0.94\columnwidth]{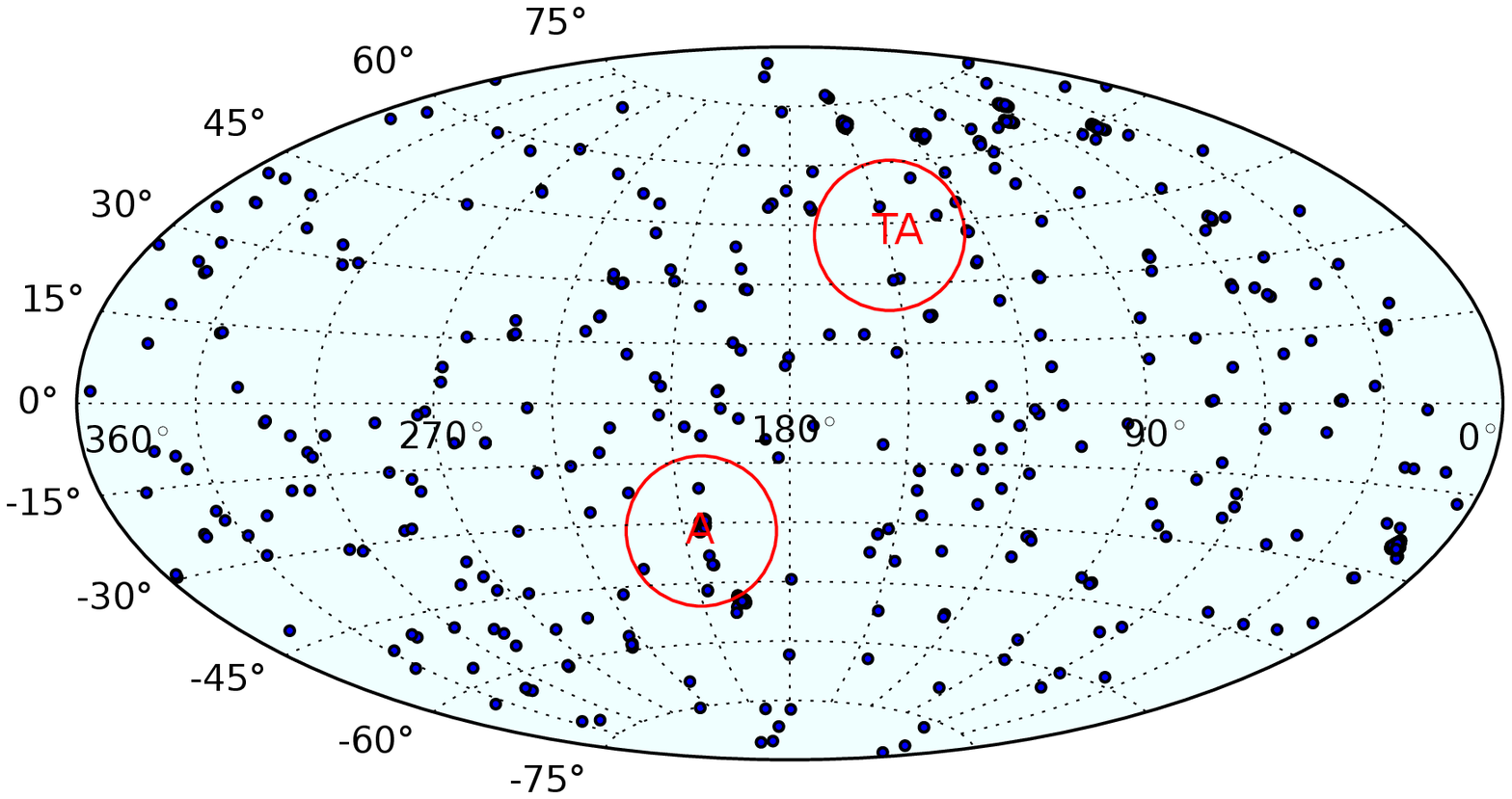}
  \caption{Aitoff projection of the sky map in equatorial coordinates of (top) the distribution of nearby starburst galaxies, the size of the marker (red circles) being proportional to the source relative FIR luminosity (weight), and (bottom) the distribution of arrival directions of simulated UHE cosmic-ray protons ($E>57$~EeV). Also shown are the locations of the TA and Auger (A) excesses \citep{ABB14, AAB15ApJ}.}
  \label{fig:Aitoff}
\end{figure}

\begin{figure} \centering
  \includegraphics[width=0.88\columnwidth]{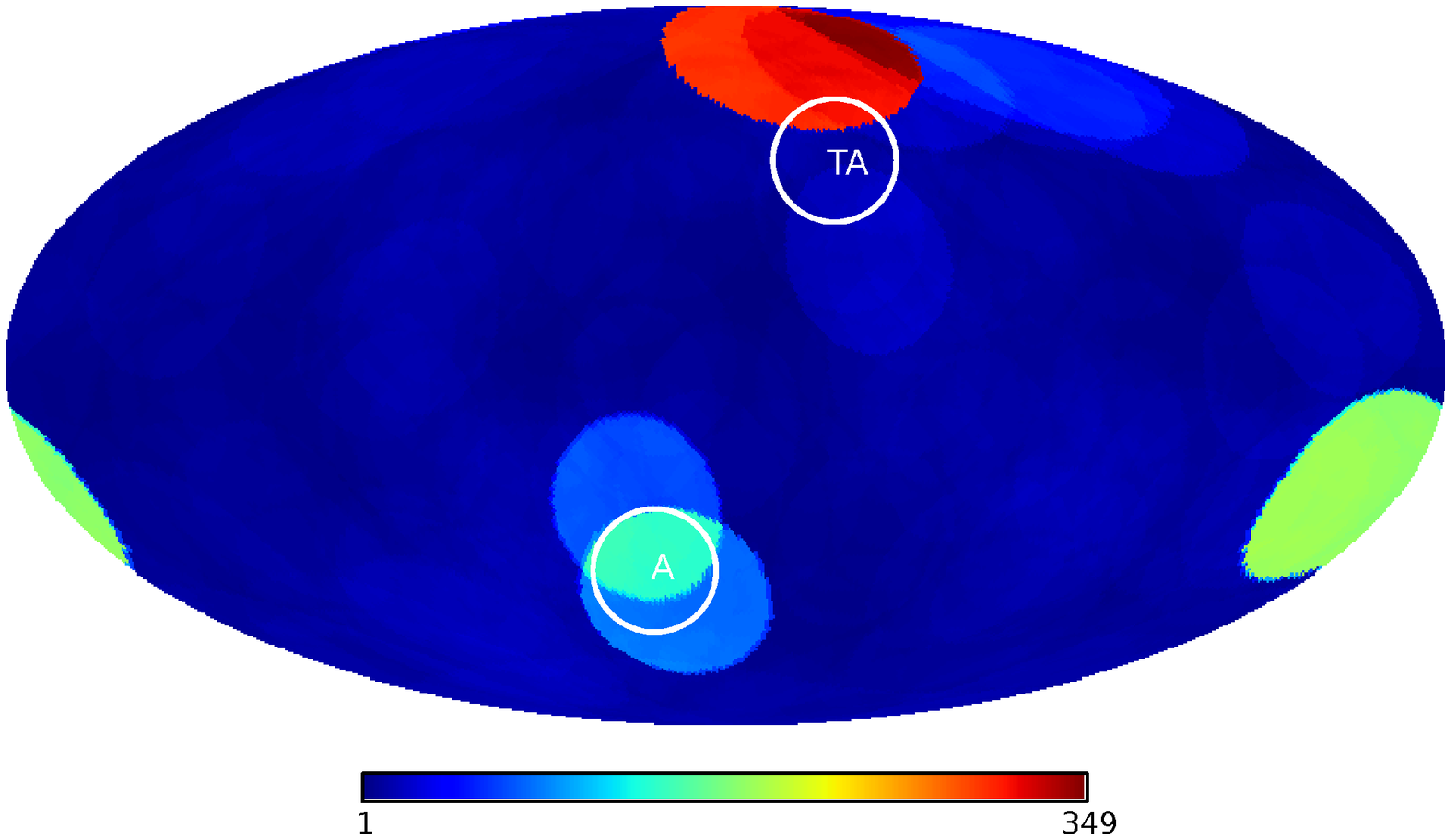}
  \includegraphics[width=0.88\columnwidth]{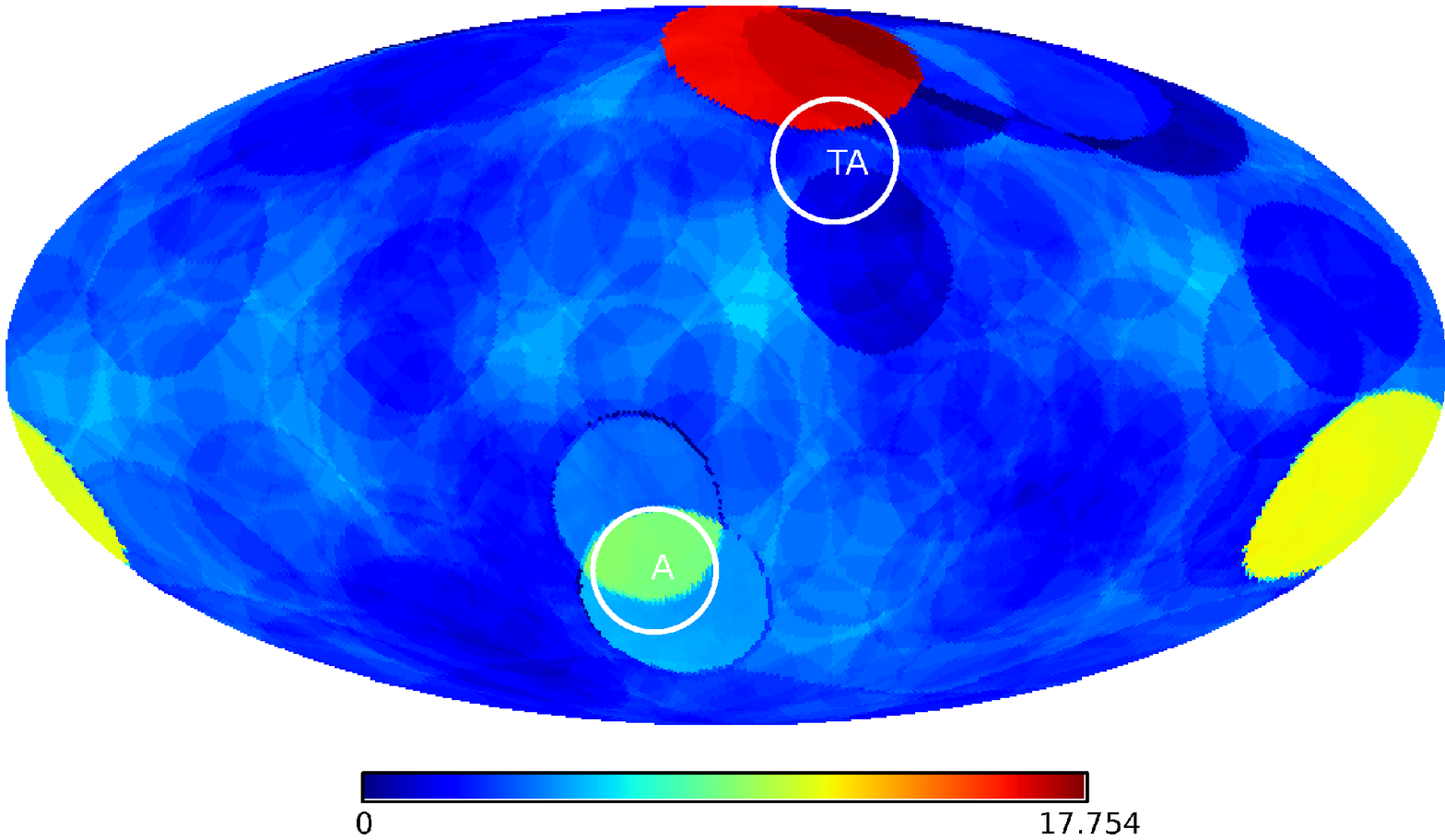}
  \caption{Mollweide projection of the sky map in equatorial coordinates of the simulated UHE cosmic-ray protons ($E > 57$~EeV) after oversampling with an angular scale of $20^\circ$ (top), and its corresponding Li-Ma statistical significance map (bottom). The color indicates the number of events per pixel (top), and the value of the statistical significance $S$ of the signal in the pixel (bottom).}
  \label{fig:mollpro1}
\end{figure}

Fig.~\ref{fig:Aitoff} shows the Aitoff projection in equatorial coordinates of the spatial distribution of the closest starburst galaxies, and the arrival direction distribution of simulated UHE cosmic rays ($E>57$~EeV). Besides M82 which might be behind the TA hotspot, the most interesting objects are NGC~4945 of the nearby Centaurus~A group, and the so-called Sculptor Galaxy (NGC~253). These three potential sources are the brightest FIR starburst galaxies in the local Universe and are all located at less than 4~Mpc from the solar system \citep{BEC09}. Fig.~\ref{fig:mollpro1} shows the Mollweide projection of the sky map in equatorial coordinates of the simulated cosmic-ray arrival directions, and its corresponding Li-Ma statistical significance map for the pure-proton mass composition. As can be seen, the model leads to several cosmic-ray excesses. The most significant one is as expected located around M82 but it doesn't perfectly coincide with the TA hotspot. This slight discrepancy may be a consequence of the IGMF model, which is most likely oversimplified. The second excess, which coincides quite well with the Auger warm spot, is not caused here by Centaurus A, as is generally believed, but rather by its close neighbor NGC~4945. Furthermore, this scenario predicts a third cosmic-ray excess (a warm spot) around the direction of the Sculptor galaxy NGC~253. This prediction can be used to test the model when statistics become sufficient. Yet, a new analysis of the dataset from the Pierre Auger Observatory has brought more credibility to this scenario, indicating a correlation between data and nearby starburst galaxies, mainly M82, NGC~4945 and NGC~253 \citep{AAB18}.

\begin{figure} \centering
  \includegraphics[width=0.88\columnwidth]{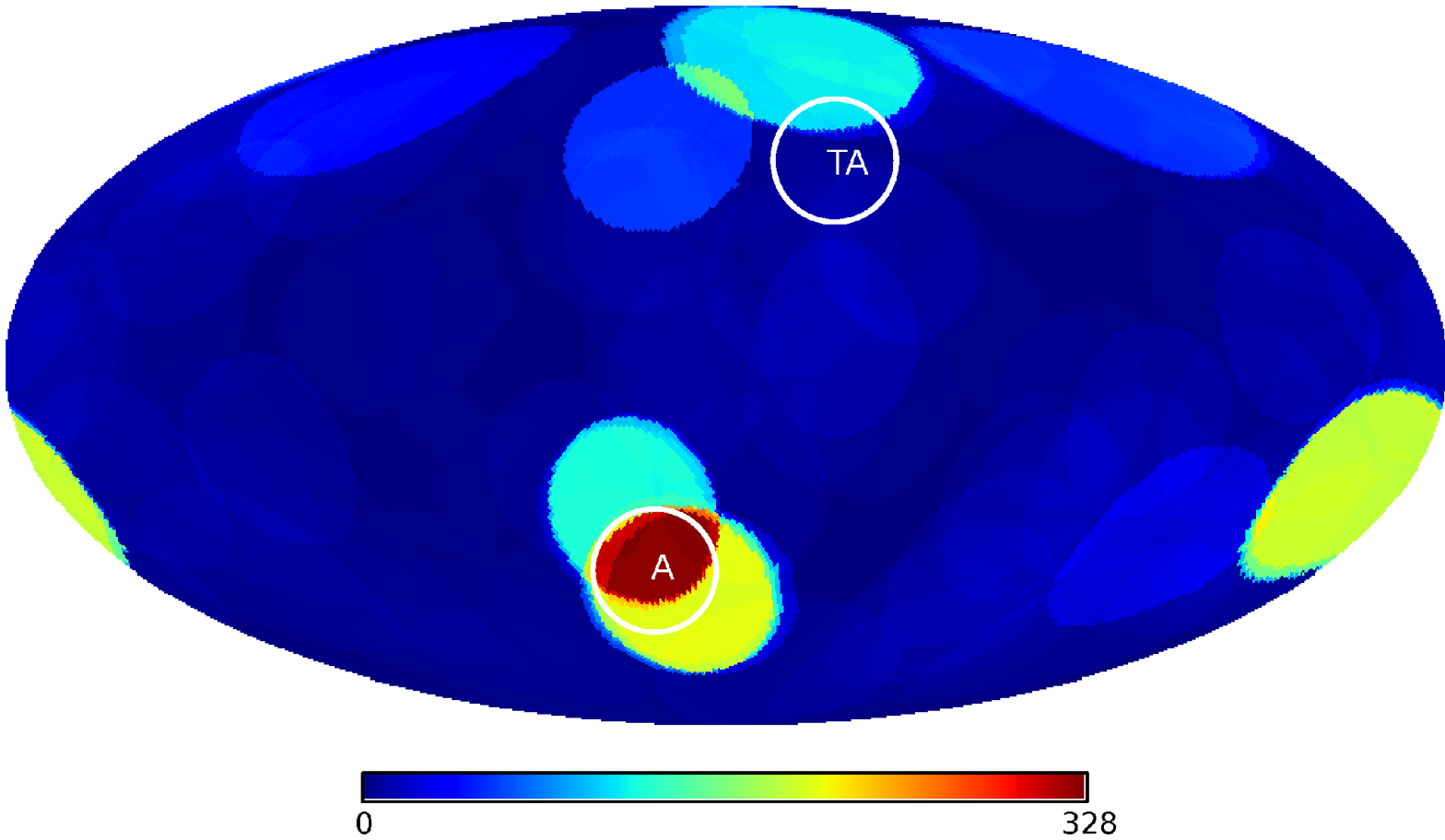}
  \includegraphics[width=0.88\columnwidth]{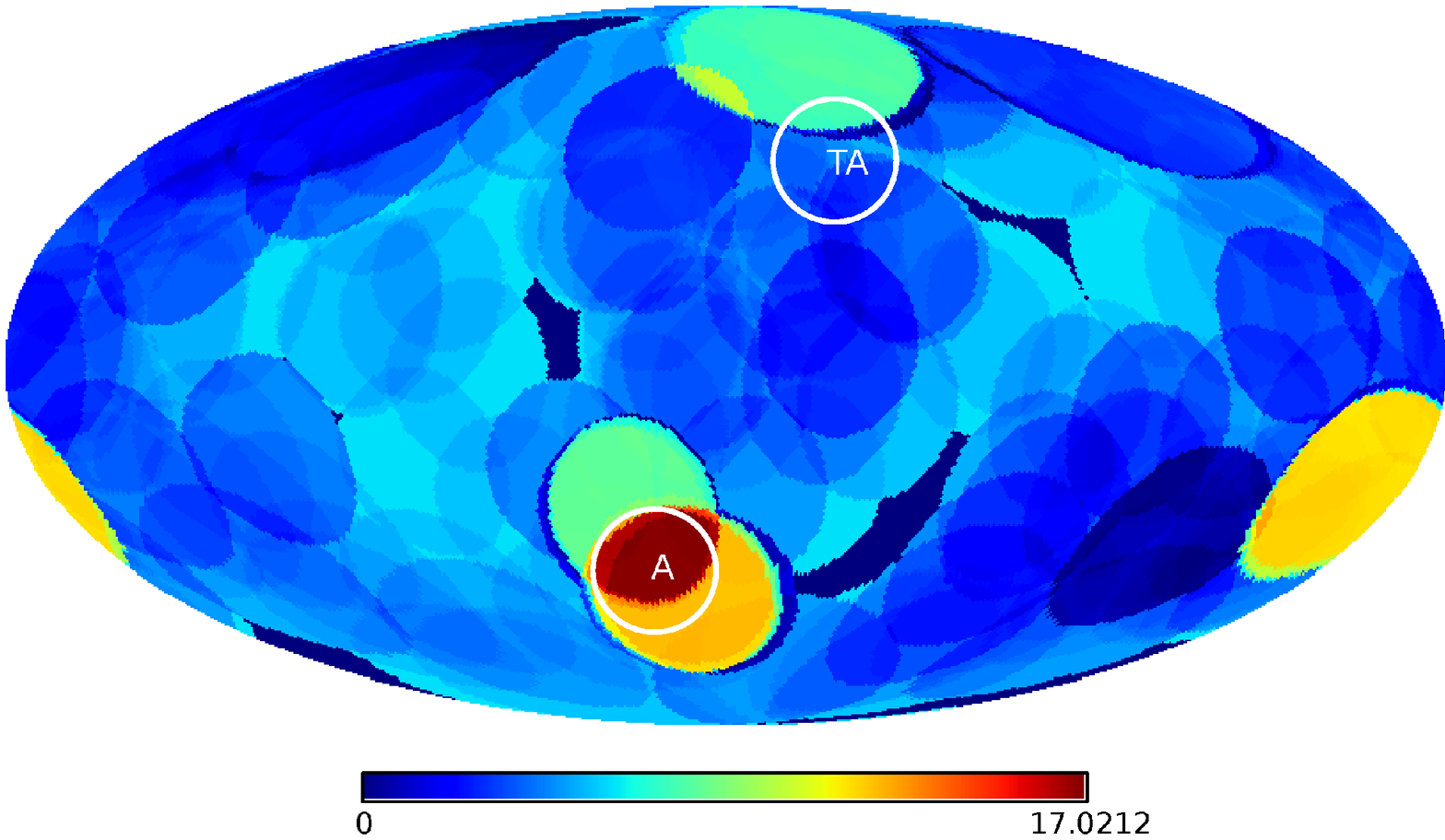}
  \caption{Mollweide projection of the sky map in equatorial coordinates for the mixed composition ($E > 57$~EeV) after oversampling with an angular scale of $20^\circ$ (top), and its corresponding Li-Ma statistical significance map (bottom).}
  \label{fig:mollmix1}
\end{figure}

\begin{figure} \centering
  \includegraphics[width=0.92\columnwidth]{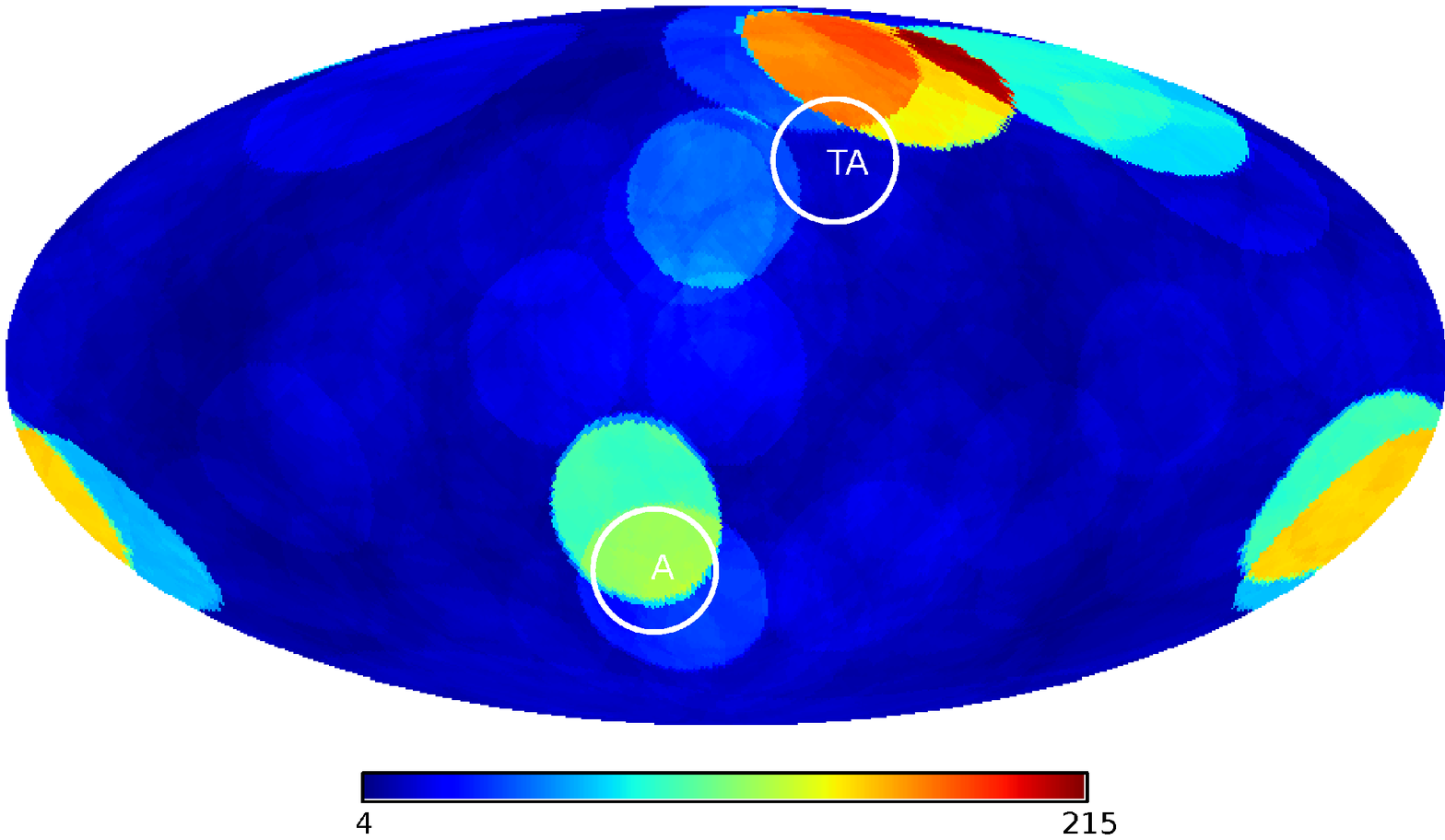}
  \includegraphics[width=0.92\columnwidth]{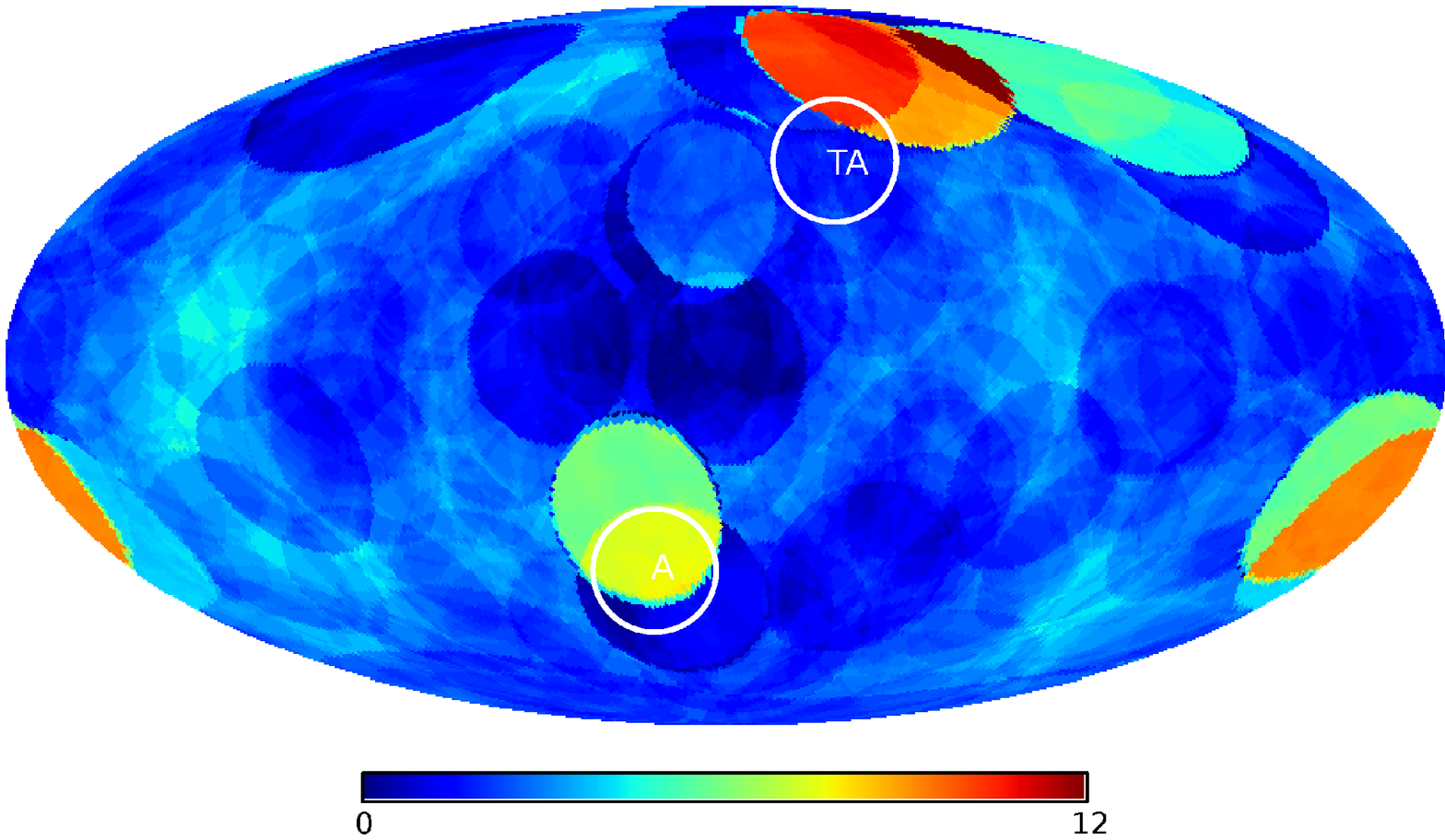}
  \caption{Mollweide projection of the sky map in equatorial coordinates for the pure-proton composition ($E > 57$~EeV) after oversampling with an angular scale of $20^\circ$ in case of an equally luminous emission (top), and its corresponding Li-Ma statistical significance map (bottom).}
  \label{fig:mollpro2}
\end{figure}

\begin{figure} \centering
  \includegraphics[width=0.92\columnwidth]{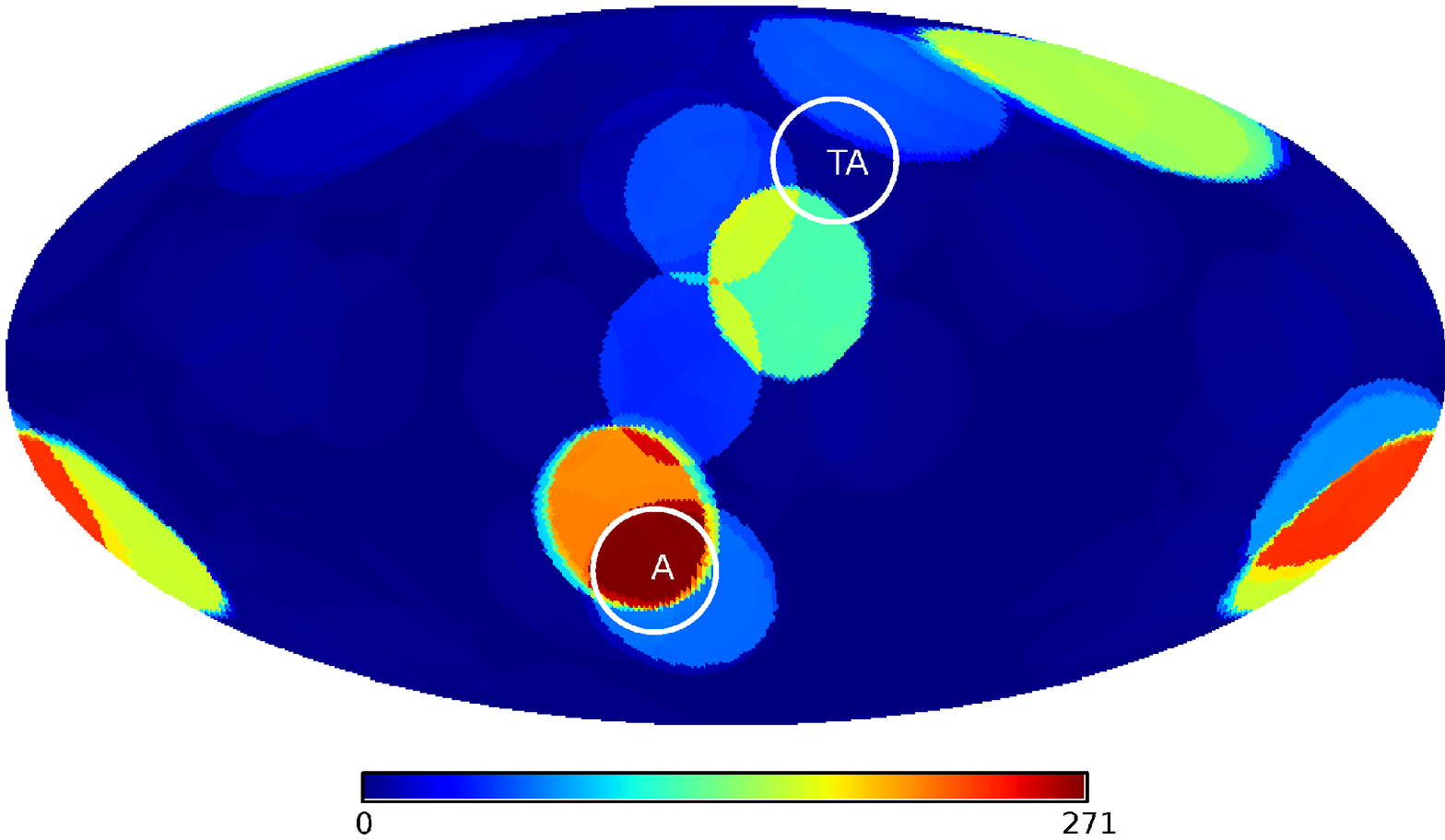}
  \includegraphics[width=0.92\columnwidth]{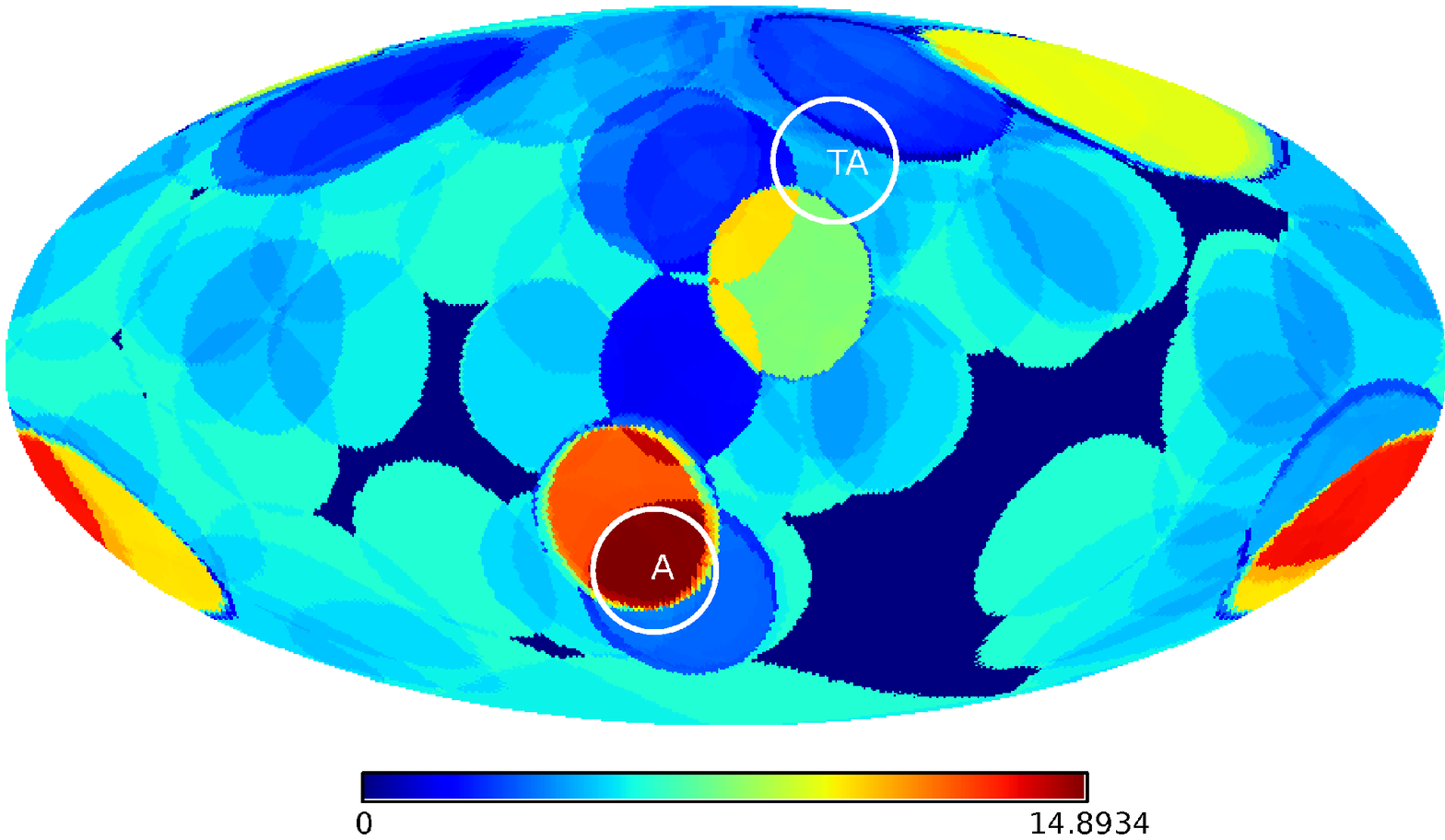}
  \caption{Mollweide projection of the sky map in equatorial coordinates for the mixed composition ($E > 57$~EeV) after oversampling with an angular scale of $20^\circ$ in case of an equally luminous emission (top), and its corresponding Li-Ma statistical significance map (bottom).}
  \label{fig:mollmix2}
\end{figure}

In case of the mixed mass composition, the obtained results are not all that different (Fig.~\ref{fig:mollmix1}). Here the three cosmic-ray excesses arise from particles coming predominantly from very close objects ($\lesssim 4$~Mpc). The deflection angle for a proton after propagation over a distance $D$ in a random magnetic field of r.m.s. strength $B$ and correlation length $\lambda$ is given by (see, e.g., \citet{STA09}):
\begin{equation}
  \langle \theta \rangle = 2.5^\circ \left( \frac{B}{1 \, \mathrm{nG}} \right) \left( \frac{D}{100 \, \mathrm{Mpc}} \right)^{1/2}
  \left( \frac{\lambda}{1 \, \mathrm{Mpc}} \right)^{1/2} \left( \frac{E}{100 \, \mathrm{EeV}} \right)^{-1}
\end{equation} 
\noindent For heavier nuclei the deflection angles are also proportional to their electric charge. For $B = 1$~nG, $D = 4$~Mpc, $\lambda = 0.23$~Mpc and $E = 60$~EeV, $\langle \theta \rangle$ is about $0.4^\circ$ for protons and $2.8^\circ$ for CNO nuclei. This small difference of only a couple of degrees explains why Fig.~\ref{fig:mollpro1} and Fig.~\ref{fig:mollmix1} are quite similar. However, there is one notable difference. Indeed, the situation is somehow reversed when compared to the case of pure-proton. The ``Auger'' cosmic-ray excess becomes the most significant at the expense of the ``TA'' one, which is not really in line with the observations.

For the sake of completeness, we also examined how the results will change if the starburst galaxies are equally luminous regarding the emission of UHE cosmic rays, rather than assuming a scaling with the FIR luminosity. We focused only on anisotropy, setting aside the all-particle energy spectrum and cosmogenic secondary particles. As seen in Fig.~\ref{fig:mollpro2} and Fig.~\ref{fig:mollmix2} which are related to the cases of pure-proton and mixed compositions, respectively, this assumption results in a bit more cosmic-ray excesses with lower statistical significance. Since only two cosmic-ray excesses have been observed so far, we are a bit further away from reality. But of course experimental statistics are still insufficient to be conclusive.

Although 3D-simulation closely reflects the physical reality, there is however one caveat to this approach. With a radius fixed to 20~kpc, the Milky Way appears as a point-like target compared to source distances. As a result, the hit probability is extremely low and, more importantly, the numerical calculations are prohibitively time-consuming. A typical run takes almost three weeks on a modern CPU running at full speed. Performing more than one realization of event injection is thus impractical, which is often the case in Monte Carlo simulations involving lengthy computing times. It is however well known that a single measurement provides only crude estimations; ideally one should make many independent realizations to make it possible to analyze errors. Nevertheless, the error in this kind of applications scales as $1/ \sqrt{N}$, $N$ being the number of simulated events, and the best is then to use the greatest possible value of $N$ to reduce errors. This is precisely the strategy we followed in this study; the value of $N$ ($=10^3$) used here is a trade-off between accuracy and computational cost.

\section{Conclusion}
\label{sec:conclusion}
The observation by the TA experiment of a cosmic-ray hotspot above 57~EeV not far from the direction of M82 logically puts forward the idea that these particles may well originate from nearby starburst galaxies. Given that there are over a hundred nearby starburst galaxies and that the (only) observed hotspot is very close to the most active of them (M82), it is not unreasonable to assume that the intensity of UHE cosmic rays from these objects somehow scales with the star formation activity, more specifically with the SFR. The simplest way to put this effect into action is to weight each source by its relative FIR luminosity observed at 60~$\mu$m, which is a good indicator of the SFR. This work clearly demonstrates that this scenario is appealing in many ways. It reproduces the all-particle energy spectrum at the highest energies, meets the test of cosmogenic particles and, on top of that, reproduces simultaneously the TA hotspot and the Auger warm spot. Regarding this last point, it is generally required the use of two different classes of sources to interpret observations \citep{BIE16}. The proximity of sources in this model is such that the primary particles do not effectively experience the GZK effect. The cutoff observed at $\sim 40$~EeV is then a sign that the sources have reached their limits with respect to cosmic-ray acceleration. On the other hand, the model does not account for the ankle structure and, in particular, suggests that the majority of cosmic-rays at $\sim 1$~EeV are not of extragalactic origin. It cannot also decide on the issue of mass composition at this point. Last but not least, the model predicts another warm spot towards the direction of the Sculptor galaxy NGC253. Such a prediction can be used to assess the plausibility of the model, but for now experimental data are still not enough to draw any firm conclusion.

It should also be noted that the results presented in this paper are based in each case on just a single Monte Carlo replication, making it impossible to assess accuracy. To reduce errors, we considered the highest possible values of the simulated events ($N$), as the error in this kind of applications scales as $1/\sqrt{N}$. The value of $N \ (= 10^{6})$ used in 1D-calculations (\S~\ref{subsec:spectrum} and \ref{subsec:cosmogenic}) is fairly large to claim that accuracy is pretty good. In 3D-calculations (\S~\ref{subsec:anisotropy}) the value of $N \ (= 10^{3})$ is not really large, but remains reasonable to maintain acceptable performances in computing times.

In addition to the FIR luminosity, the gamma-ray luminosity is also a good indicator of the activity of astrophysical objects at very high energy. In the same vein, this quantity can be used to weight the potential sources of UHE cosmic rays. In our future refinements we will try to pursue this matter further, particularly in respect of anisotropy.

\section*{Acknowledgments}
We gratefully acknowledge the CRPropa team for making public their excellent simulation code, and for their valuable assistance and support. We also wish to express our thanks to the reviewers for their relevant comments, which helped substantially improve the paper.

\bibliographystyle{mnras}
\bibliography{uhecr}

\bsp
\label{lastpage}
\end{document}